# A pH-Sensor Scaffold for Mapping Spatiotemporal Gradients in Three-Dimensional *In Vitro* Tumour Models

Riccardo Rizzo,[1,§,*] Valentina Onesto,[1,§] Stefania Forciniti,[1] Anil Chandra,[1] Saumya Prasad,[1] Helena Iuele,[1] Francesco Colella,[1] Giuseppe Gigli,[1,2] Loretta L. del Mercato[1,*]

1 Institute of Nanotechnology, National Research Council (CNR-NANOTEC), c/o Campus Ecotekne, via Monteroni, 73100, Lecce, Italy.
2 Department of Mathematics and Physics ''Ennio De Giorgi'', University of Salento, via Arnesano, 73100, Lecce, Italy.

§: Equally contributing authors
*: corresponding authors

**Key words:** Optical pH-sensors; alginate microgels; *in vitro* 3D tumour models; tumour microenvironment; cell metabolism; computational analysis.

## Abstract

The detection of extracellular pH at single cell resolution is challenging and requires advanced sensibility. Sensing pH at a high spatial and temporal resolution might provide crucial information in understanding the role of pH and its fluctuations in a wide range of physio-pathological cellular processes, including cancer. Here, a method to embed silica-based fluorescent pH sensors into alginate-based three-dimensional (3D) microgels tumour models, coupled with a computational method for fine data analysis, is presented. By means of confocal laser scanning microscopy, live-cell time-lapse imaging of 3D alginate microgels was performed and the extracellular pH metabolic variations were monitored in both *in vitro* 3D mono- and 3D co-cultures of tumour and stromal pancreatic cells. The results show that the extracellular pH is cell line-specific and time- dependent. Moreover, differences in pH were also detected between 3D monocultures *versus* 3D co-cultures, thus suggesting the existence of a metabolic crosstalk between tumour and stromal cells. In conclusion, the system has the potential to image multiple live cells types in a 3D environment and to decipher in real-time their pH metabolic interplay under controlled experimental conditions, thus being also a suitable platform for drug screening and personalized medicine.

## 1. Introduction

Elevated hydrogen ion concentration accumulates in the tumour microenvironment as a result of increased cellular metabolic demand and altered perfusion, e.g., oxygen availability or acidic metabolic waste products.(Boedtkjer et al. 2012) An important contribution to the extracellular acidification is provided by metabolic reprogramming of cancer cells towards increased aerobic glycolysis and reduced mitochondrial oxidative phosphorylation, referred to as Warburg effect,(Schwartz et al. 2017) which is a typical feature of cancer. This metabolic reprogramming confers to cancer cells protection against oxidative stress, enhances their resistance to hypoxia and determines a massive glucose uptake.(Swietach et al. 2014) As a consequence, the functional interplay between cancer and stromal cells, as well as their interactions with the extracellular matrix are profoundly affected. Overall, the acidity of the tumour microenvironment, which is spatially and temporally heterogeneous,(Helmlinger et al. 1997; Rohani et al. 2019) affects cancer initiation and progression,(Boedtkjer and Pedersen 2020) but also the efficacy of anti-cancer drugs treatments.(Tredan et al. 2007) Therefore, monitoring the local pH metabolic fluctuations is critical in understanding the basic biology of the tumour, and can also be used as a valid metabolic readout for cancer diagnosis and treatment.(Parks et al. 2011; Schwartz et al. 2017) Importantly, the complexity of the tumour microenvironment, which includes cell-cell interactions and extracellular matrix composition, coupled with the fast diffusion and mobility of extracellular protons, makes the extracellular pH mapping extremely challenging.(Parks et al. 2011; Piper et al. 2008)

In acid-base regulation studies, therefore, there is an urgent need to generate reliable tools for spatio-temporal pH detection, and a strong demand for stable and biocompatible *in vitro* systems that mimic the geometry and structural composition of solid tumours.(Boedtkjer and Pedersen 2020; Moldero et al. 2020) Different methods have been developed to monitor pH in living cells, ranging from conventional microelectrodes(Nadappuram et al. 2013) to more sophisticated label-free pH-sensitive nanoprobes,(Zhang et al. 2019) which both rely on the movement of one or multiple electrodes in different positions, thus limiting the spatio-temporal resolution. Moreover, these systems are more invasive than optical approaches.(Anemone et al. 2019) Among the latter, ratiometric fluorescence-based pH sensors have been proved to be highly suitable for real-time and *in situ* analyses at cellular level thanks to their minimal invasive features and high reliability in terms of measurements, which are independent from probes concentration changes, instrument sensitivity and environmental conditions.(Chandra et al. 2021; De Luca et al. 2015; del Mercato et al. 2015; Lee et al. 2015; Lee and Kopelman 2012; Moldero et al. 2020; Naciri et al. 2008; Nielsen et al. 2010; Prasad et al. 2021; Roy et al. 2020; Schaferling 2012; Shi et al. 2012; Steinegger et al. 2020; Wencel et al. 2014; Xue et al. 2017; Yang et al. 2018) In ratiometric pH sensing, together with the pH indicator fluorophore, an additional pH insensitive fluorophore is used as a reference, and ratio of their emission intensity is correlated with pH change.(Chandra et al. 2022) Thus, a specific emission ratio will indicate a specific pH value. This makes the pH sensing more accurate, as ratio of fluorescence intensities nullifies the effect of concentration variation that may mislead. For example, using only FITC to prepare the silica microparticle pH sensors, would have resulted in fluorescence emission due to both the pH change and change in number of sensors. Thus, in the current work ratiometric sensors are developed using both FITC and RBITC. In addition to fluorescence-based pH sensors there are now newer reports where other techniques like surface-enhanced Raman spectroscopy (SERS) further extended the application. This was achieved by designing dual-signal optical sensor particles for intracellular and extracellular pH imaging.(Pallaoro et al. 2010; Wiercigroch et al. 2019; Yang et al. 2019)

As a model system to study the pH associated phenomena in tumors, *in vitro* 3D cancer cells models are proven to be suitable platforms, since they can recapitulate most of the tumour microenvironmental features that are observed in solid tumours, including signalling factors and metabolic gradients.(Cavo et al. 2020; Delle Cave et al. 2021; Friedrich et al. 2009; Gjorevski et al. 2016; Hakanson et al. 2011; Langer et al. 2019; Nath and Devi 2016; Osuna de la Pena et al. 2021; Rijal and Li 2017; Turetta et al. 2018; Worthington et al. 2015) In the current work, to realise a 3D cell culture platform enabled with ratiometric pH sensing capabilities, we developed a spherical *in vitro* 3D cell culture and microparticle based pH-sensor platform that is compatible with confocal laser scanning microscopy (CLSM), allowing visualization and detection of acid-base metabolic variation at single cell level. Notably, the generation of spherically structured scaffolds allows a fine control on the biophysical properties of the produced spheroids, and guarantees an easy reproducibility of the shape and size of the hydrogels, key aspects in drug treatment and screening.(Ferreira et al. 2018) It is well known that acidic microenvironment is common in solid tumours, particularly in human pancreatic adenocarcinoma (PDAC), one of the most aggressive and intractable cancers.(Ho et al. 2020; Siegel et al. 2021) A crucial hallmark of PDAC is the extensive desmoplastic stroma that accounts for the 90% of the total tumour mass and that is considered originating primarily from activated pancreatic stellate cells (PSC).(Biffi et al. 2019; Ohlund et al. 2017) During tumorigenesis, PSCs change into an active myofibroblast-like phenotype and start depositing large amounts of extracellular matrix, creating a favourable and supportive acidic tumour microenvironment that facilitates cancer progression and metastasis.(Hwang et al.

2008; Xu et al. 2010) For these reasons, human pancreatic adenocarcinoma (PDAC) cell lines and pancreatic stellate cells (PSCs) were selected as cellular model system. The realized *in vitro* pH-sensor 3D platform is obtained through a multistep fabrication procedure including (i) synthesis and characterization of silica-based fluorescent ratiometric pH sensors; (ii) integration of pH sensors within biocompatible 3D alginate microgels in presence of stromal and cancer cells, either alone or in combination; (iii) CLSM time-lapse imaging and quantification over time and space of single cell extracellular pH metabolic variations.

## 2. Results and discussion

### 2.1 Generating spherical 3D culture platform for extracellular pH sensing

To generate spherical 3D *in vitro* models for extracellular pH evaluation, we have initially synthetized silica-based fluorescent pH sensors as described in methods (see also **Figure S1**). To this aim, the fluorescein isothiocyanate (FITC) and the rhodamine B isothiocyanate (RBITC) dyes were selected as pH-sensitive and pH-insensitive dyes (Han and Burgess 2010; Le Guern et al. 2020; Lei et al. 2010), respectively, to create fluorescent $SiO_2$ microparticles (mean particle diameter 1.521 $\pm$ 0.073 µm) for ratiometric pH sensing. The UV-Vis absorption and pH dependent emission spectrum of the sensors is shown in **Figure S2a** and **b**, where the fluorescence plot indicates sensitivity of FITC and insensitivity of RBITC molecules, thus making them a suitable ratiometric pair for pH sensing. To confirm the presence of negative charge on the microparticles, zeta potential estimation was done and found to be -84 mV which is in agreement with typical values for $SiO_2$-based nano- and microparticles (**Figure S2c**). The spherical morphology of the $SiO_2$ microparticles was confirmed using scanning electron microscopy (SEM), which showed presence of highly monodispersed solid $SiO_2$ microparticles (**Figure S2d**). Next, the response of sensors to pH-adjusted cell media was evaluated in detail by means of fluorimeter and CLSM measurements, respectively (**Figures 1a** and **c**). The sensors show a linear response in the range between pH 5.0 and pH 8.0, with a correlation coefficient $R^2$ = 0,976 (**Figure 1a**). **Figure 1c** shows representative CLSM images of $SiO_2$ pH sensors exposed to 5.0, 6.0, 7.0 and 8.0 pHs. The emission of individual sensors strongly depends on the local pH, where the $SiO_2$ microparticles at neutral and basic pH strongly emit the green light by FITC compared to the intensity of red light by RBITC. At acidic pH the contribution of green emission from FITC gets significantly reduced while the red emission from RBITC remains unchanged, thus the resultant overall color of the particles in the merged image appears red-orange indicating local acidic pH. Thus the recorded color change of the particles is only taking place due to FITC molecule because of its unique pH sensitive photophysical properties of the used pH indicator dye, FITC.(Le Guern et al. 2020; Lei et al. 2010) Moreover, the microparticles sensors show a good reversibility to pH switches and a good stability over time (**Figure 1b**), indicating their high suitability for time lapse pH sensing analyses in biological systems.

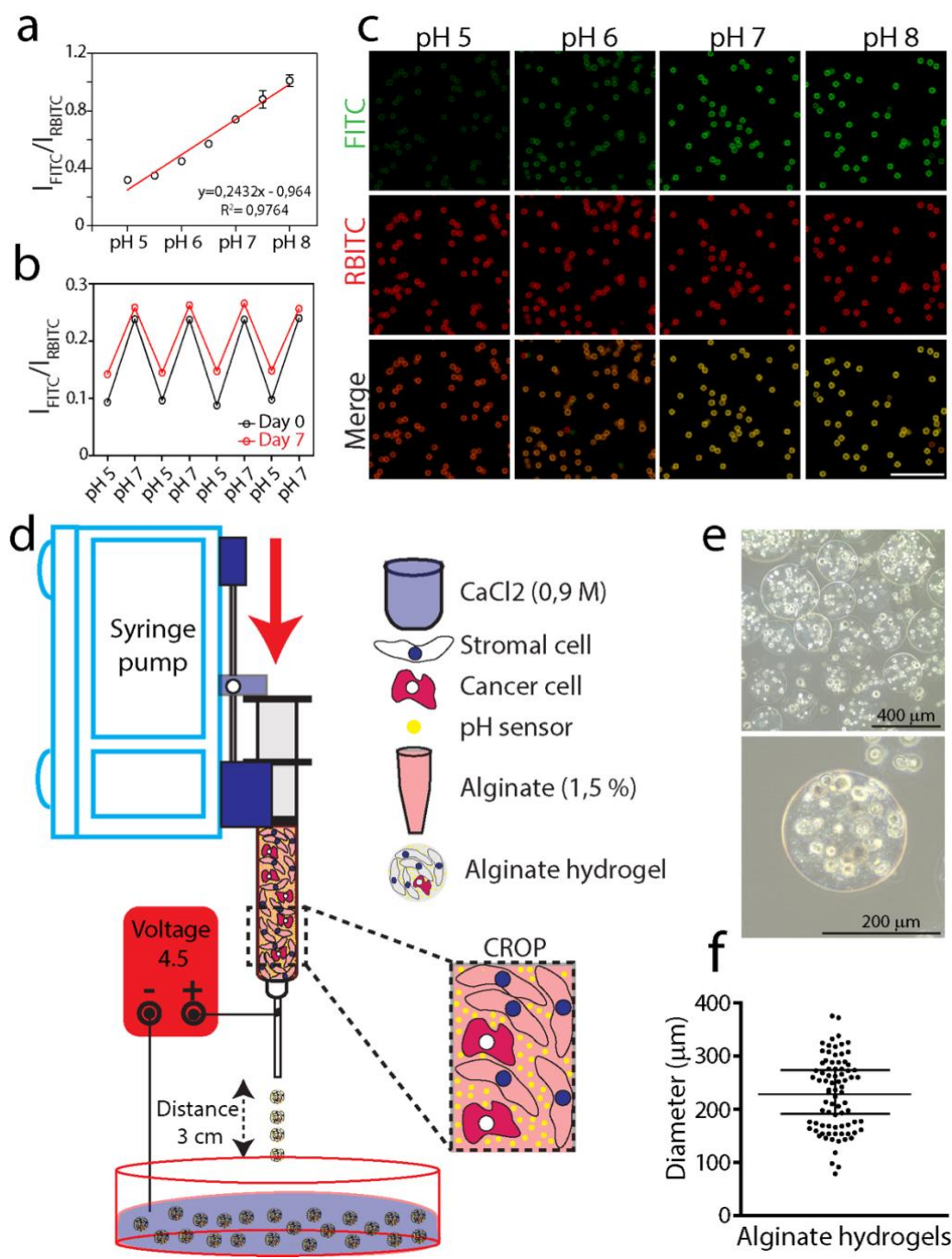

**Figure 1**. **Generation of 3D co-culture system for extracellular spatio-temporal pH sensing**. **(a)** pH calibration curve of SiO$_2$-based sensors as a function of pH: the fluorescence intensity ratio between the pH sensitive fluorophore FITC at 500-700 nm ($\lambda_{exc}$: 488 nm) and the reference fluorophore RBITC at 565-700 nm ($\lambda_{exc}$: 561 nm) was plotted *vs*. the pH of the pH-adjusted cell media. Each plotted value corresponds to the mean value of 3 independent readings. Error bars show the standard deviation. **(b)** Reversibility and ageing of SiO$_2$-based pH sensors: the fluorescence intensity ratio of FITC and RBITC signals was recorded by series of pH switches between pH 5 and pH 7 (n=3). The measurements were performed after sample preparation (day 0, black line) and repeated after 7 days (day 7, red line). **(c)** Representative CLSM micrographs showing the pH-dependence of microparticles fluorescence in pH-adjusted cell medium (FITC was excited at 488 nm and RBITC at 561 nm). Green channel (500-550 nm), red channel (570-620 nm), and the overlay of the fluorescence channels are reported. Scale bars, 20 µm. **(d)** Schematic set up of microencapsulation system for the generation of spherical alginate hydrogels containing cancer cells, stromal cells and pH-sensor microparticles. Objects are not drawn to scale. **(e)** Representative optical micrographs showing the shape and the size of alginate microgels containing AsPC-1 cells, PSCs cells and pH-sensors. **(f)** Average diameter of alginate microgels shown in (**e**) (n=3). Individual data points are shown in the graphs; data are means ± SEM of at least three independent experiments.

Then, we developed a lab-made high voltage microdroplet encapsulation apparatus and we optimized a protocol for cells and pH sensors microencapsulation into spherical 3D alginate microgels. The rationale for selecting alginate is its biocompatibility and its transparency that makes it highly suitable for optical microscopy,(Andersen et al. 2015) i.e. CLSM live cell imaging (see methods and scheme in **Figure 1d**). Different key parameters affecting hydrogel's composition and morphology(Gansau et al. 2018) (i.e. alginate concentration, emitter needle gauge, applied voltage, flow rate and cell seeding density) were tested and the most optimal configuration for 3D time-lapse imaging and pH sensing was selected. Optical images reported in **Figure 1e** show the morphology of as-obtained hydrogels, which appear smooth and spherical with a mean diameter of 228.4 ± 39.80 µm (SEM, **Figure 1f**).

Hydrogels are relatively homogeneous and spherical in shape (**Figure S3a**) with a regular porous polymer network entrapping cells and pH sensors (**Figures S3b-c**). The morphology of pH sensors was further studied by Transmission Electron Microscopy (TEM) (**Figures S3d-f**), which confirmed their spherical shape and a mean diameter of 1.299±0.017. To note, average diameters extracted from TEM images were in accordance with the values reported by DLS analysis (see experimental section). However, because of the known dynamic swelling in water, average hydrodynamic sizes resulted being slightly bigger than the ones evaluated under dry condition.(Kersting et al. 2020) Importantly, samples with diameter below 200 µm, which were separated with pluriStrainer® (200 µm mesh sizes), were selected for extracellular pH mapping by CLSM, since their size was suitable for the selected objective (see experimental section for further details) and effective penetration of the light along the z-axis, thus providing a good resolution for image processing and analysis.(Jonkman et al. 2020)

## 2.2 Scaffold Biocompatibility

Next, we tested the scaffold's biocompatibility by evaluating two parameters: i) the diffusion of nutrients within the whole hydrogel and ii) cell viability. To address the first question, PSCs stromal cells, AsPC1 or PANC-1 tumour cells were microencapsulated into the spherical 3D alginate scaffolds. Then, glucose uptake by cells was visualized by fluorescence microscopy (**Figure 2a**) and quantified over time by flow cytometry (**Figure 2b**) using the fluorescent glucose analogue 2-NBDG (2-[N-(7-nitrobenz-2-oxa-1,3-diazol-4-yl) amino]-2-deoxy-D-glucose), as described in methods. Cells which uptake 2-NBDG in the hydrogel are green (**Figure 2a**), thus confirming that nutrients diffuse within the 3D scaffold. Indeed, flow cytometry results show that the glycolytic demand of cells encapsulated in alginate hydrogels is cell line-specific and time-dependent (**Figure 2b**). These data support the use of spherical alginate microgels as a 3D co-culture system, where two or more cell types can exchange nutrients and metabolites, thus providing a good assay to investigate intercellular metabolic interplay.

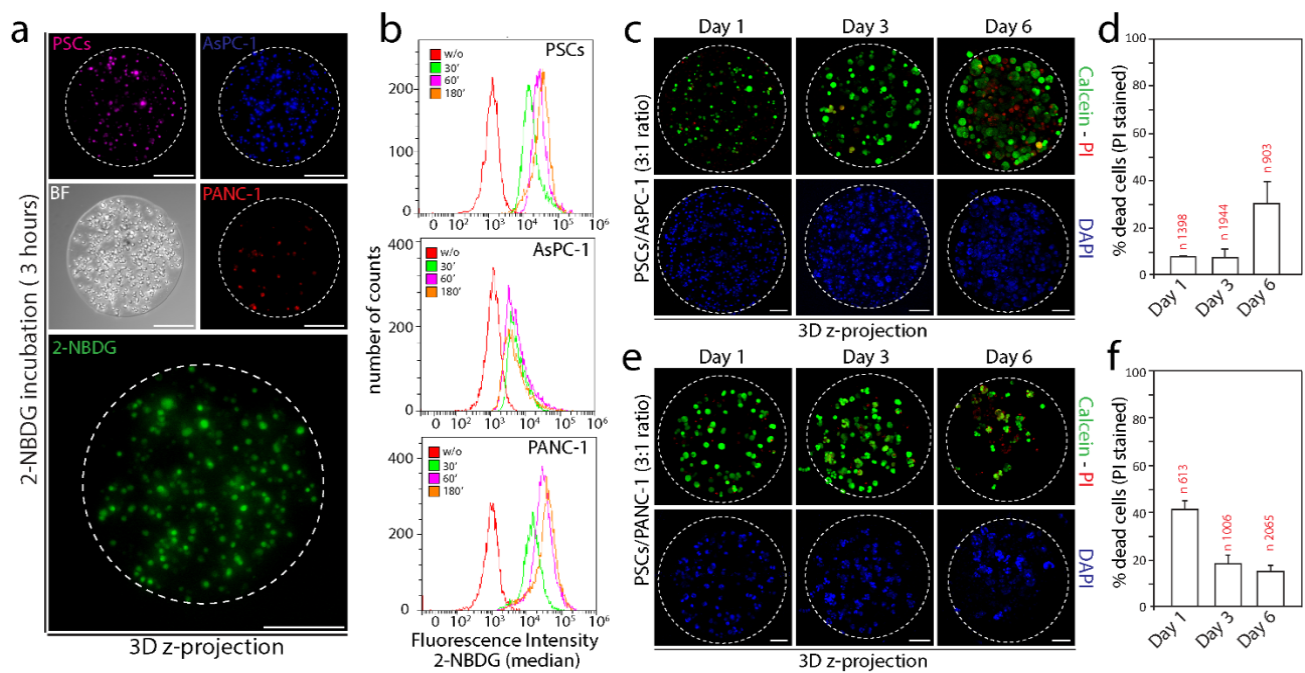

**Figure 2. Diffusion of nutrients and cell viability in spherical alginate hydrogels**. **(a)** Visualization of cellular uptake of fluorescent 2-NBDG in spherical alginate hydrogels: representative micrographs showing alginate hydrogel (maximum intensity projection) containing cancer and stromal cells following exposure to 2-NBDG (100 μM) for 3 hours. AsPC-1 (Hoechst, blue), PANC-1 (RCA I Rhodamine, red); PSCs (Deep Red, magenta), bright field (BF, grey). Cells which uptake glucose are in green (2-NBDG). Dashed lines indicate the edge of the hydrogels. Z-projections of 76 sections; Z-stack step size = 5 μm. Scale bars, 200 μm. **(b)** Flow cytometry analysis of cellular uptake of fluorescent 2-NBDG in spherical alginate hydrogels. Hydrogels containing AsPC-1 or PANC-1 or PSCs were treated with fluorescent 2-NBDG (100 μM) at each indicated time point, then the cells were released from alginate and analysed by flow cytometry. **(c-f)** Cell viability in alginate hydrogels. **c, e)** Fluorescence microscopy images (maximum intensity projection) of PSCs and AsPC-1 (**c**) or PANC-1 (**e**) cells encapsulated in alginate hydrogels (3:1 ratio) and stained at different days with calcein (green, live cells), propidium iodide (red, dead cells) and Hoechst (nuclei, blue). Dashed lines indicate the edge of the hydrogels. Scale bars, 100 μm. In **c**) Z-projections of 87 sections for day 1; 114 sections for day 3; 66 sections for day 6. In **e**) Z-projections of 85 sections for day 1; 71 sections for day 3; 72 sections for day 6. Z-stack step size = 2.55 μm. **d)** Quantification of the experiment in **c**) and **e**) is plotted in **d**) and **f**), respectively (n indicate the number of cells). Data are means ± SEM.

Subsequently, we have evaluated the scaffold toxicity. To this aim, PSCs stromal cells were microencapsulated with AsPC1 or PANC-1 cancer cells in 3:1 ratio (to mimic the ratio found in physiological tumours(Frantz et al. 2010; Neesse et al. 2015) in the presence of $SiO_2$ microparticles, and cell viability as measured by calcein-AM (green) and propidium iodide (PI, red), was monitored by CLSM for several days (**Figures 2 c** and **e**). As noted by nuclear staining, cells grow and divide from day 1 to day 6, thus moving from single spread entities (day 1) to a cluster of cells (cell spheroids, days 3-6), which occupied the empty space of the hydrogel over time (**Figures 2 c** and **e**). In both PSCs/ AsPC-1 and PSCs/PANC-1 3D co-culture systems, most of the cells show green fluorescence (**Figures 2 c** and **e**). However, the number of dead cells (red), which were quantified at different time points and expressed as a percentage of the total (blue, Hoechst), indicates higher cell viability in the case of PSCs/AsPC-1 respect to PSCs/PANC-1 3D co-culture systems (**Figures 2 d** and **f**). Basing on these results, we decided to use PSCs and AsPC-1 cells as a co-culture system to be coupled with alginate-based hydrogels, thus generating a suitable biocompatible spherical 3D culture platform for spatio-temporal extracellular pH sensing.

## 2.3 Algorithm designed for 3D time lapse processing and analysis of pH sensors

The pH response of $SiO_2$-based pH sensors loaded within 3D alginate hydrogel was analysed in details by a customized algorithm designed for the automatic detection of both pH sensors and cells, and for the extraction of the pH read-outs. The code pipeline is shown in **Figure 3** and described in Experimental Section. The algorithm takes as input z-stack and time-lapse CLSM records (**Figure 3a**), with four fluorescent channels, corresponding to the pH-sensitive fluorophore FITC, the reference fluorophore RBITC, the tumour cells (Deep Red) and the stromal cells (Hoechst). In case of single cell cultures, the input signals were limited to FITC, RBITC and Hoechst. Instead, during the calibration step, performed before each time-lapse CLSM acquisition, only the fluorescent signals belonging to the microparticles (i.e., FITC and RBITC) were recorded as input. The results of the calibration analyses are shown in **Figure 4** and described in detail in Experimental section. It is worth mentioning that small temperature fluctuations might affect the live imaging, thus causing shift of the objects (Jonkman et al. 2020) during CLSM acquisitions, hence, when merging FITC and RBITC channels, some of the sensors did not overlay in the (x, y) direction. Therefore, the algorithm applied a motion correction reporting sensors that were shifted in the RBITC channel to their corresponding position in the FITC channel (**Figure 3b**). Finally, the algorithm segmented (x, y, z) CLSM records (**Figure 3c**) for each time point, extracted the positions of cells and microparticles (**Figure 3d**) and calculated from FITC/RBITC ratiometric intensity measures (**Figure 2e**) pH at different spatial locations (**Figure 3g**). The code addresses the necessity of extracting both global mean pH measures both local measures, automatically selecting a cell and considering the sole sensors belonging to that specific cell microenvironment, excluding probes far from the considered cell (**Figure 3f**). The final output consists in the generation of a colorimetric spatio-temporal map of pH within a 3D system. Indeed, mapping of pH in 3D conditions is still in its infancy phase, and there are only few reports that propose a coupled approach involving pH measurements across 3D cultures together with a computational tool for the analysis of 3D time-lapse recording.(Kang et al. 2021; Zhang et al. 2021; Zhang et al. 2020) The only work presenting a mixed population of cells and sensors, does not report 3D time-lapse measurements and, moreover, pH detected from the sensors is analysed manually without an advanced automated computational tool.(Moldero et al. 2020)

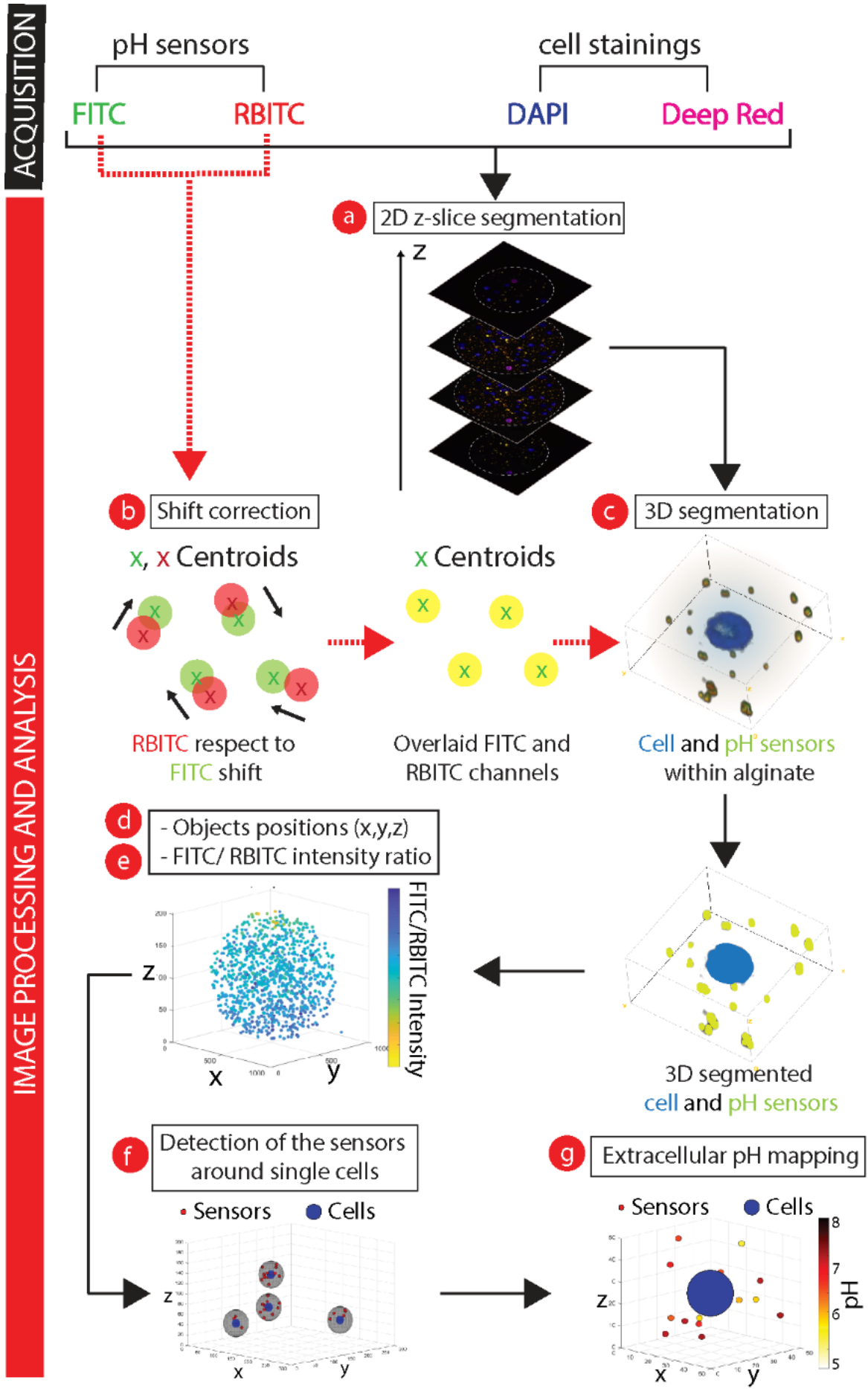

**Figure 3. Schematic representation of the workflow for the analysis and processing of 3D time lapse data set**. Red dashed arrows indicate lines of the code executed for the sole FITC and RBITC channels, the black arrows for all the input channels. Acquired data are first z-by-z segmented (**a**) to detect cells and sensors. The RBITC respect to FITC shift is corrected (**b**) and cells and sensors are reconstructed in 3D (**c**). Next, the centroids of cells and sensors are calculated (**d**) as well as FITC/RBITC ratiometric intensity measures ($I_{FITC}/I_{RBITC}$) (**e**). The sole sensors surrounding the cells are detected (**f**) and, finally, by passing $I_{FITC}/I_{RBITC}$ values to a calibration curve, spatio-temporal pH maps are recreated (**g**).

## 2.4 pH-sensing properties of spherical 3D alginate hydrogels

The pH response of $SiO_2$-based pH sensors was tested within the 3D hydrogels. For this purpose, the sensors were integrated in 3D spherical alginate hydrogel co-cultures and exposed to different pH-adjusted cell media (pHs 4.0, 4.5, 5, 5.5, 6, 6.5, 7, 7.5, 8.0). Samples were left to equilibrate for 30 minutes before being imaged, along the z-axis, by CLSM. **Figure 4a** shows representative images of spherical hydrogels incubated at pH 4.0, 5.0, 6.0 and 7.0, respectively. As expected, the fluorescence intensity of FITC (pH indicator dye) increases with pH, while the fluorescence intensity RBITC remains stable (reference signal). The intensity ratio ($I_{FITC}/I_{RBITC}$) of the sensors was extrapolated for each pH and their mean plotted as a function of pH, thus showing a direct relationship between the fluorescence intensity of the sensors and the proton concentration (**Figure 4b**). Notably, the calibration curve was setup at the beginning of each experiment and the formula of the fit calibration curve was calculated and used to measure unknown pH of sensors located within the hydrogel under evaluation. pH sensors in **Figure 4a** are also presented as 3D scatter plot in **Figure 4c**. Here, sensors show similar pH values within the same hydrogel, as it can be noticed from the pH colormap obtained from the pH calibration formula reported in **Figure 4b**.

Next, we tested the sensing stability of pH sensors in the hydrogel, at fixed pH, during 4D time-lapse confocal microscopy. After hydrogels production, the calibration curve was setup as described before, and the hydrogel under evaluation was incubated with pH-adjusted cell media (pH 7.6) and placed under CLSM in humidified atmosphere of 5% $CO_2$ and controlled temperature (see methods). The whole sample was imaged a time 0 and after 10 hours and the fluorescent emission of FITC and RBITC was recorded for each sensor. Representative images of the hydrogel are reported in **Figure S4a.** The pH value of each sensor was extrapolated based on the calibration curve and represented as 3D scatter plot (**Figure S4b**), while their mean was calculated and represented in **Figure S4c.** The obtained results confirm that the sensors remain stable over time. Notably, a minimal variation of the pH cell medium was detected, as confirmed by the post-acquisition measurements carried out by means of a pH meter (pH 7.56), thus indicating that any pH variation observed in the hydrogel could only be ascribed to proton release by the cells. To assess whether the density of the sensors affects pH detection, we performed calibration experiments on hydrogels containing increasing concentrations of pH sensors (**Figure S4d-f**). Specifically, samples were incubated with pH-adjusted cell media (pHs 4.0, 5.0, 6.0, 7.0) and the ratiometric calibration curves were extracted (**Figure S4d-f**). In all the tested conditions, calibration curves were consistent with $R^2$ values near to 1, thus demonstrating that the number of pH sensors did not affect pH detection.

Overall, these data strongly support the suitability of this system for extracellular pH evaluation over time and space at single cell level in 3D cell culture environment.

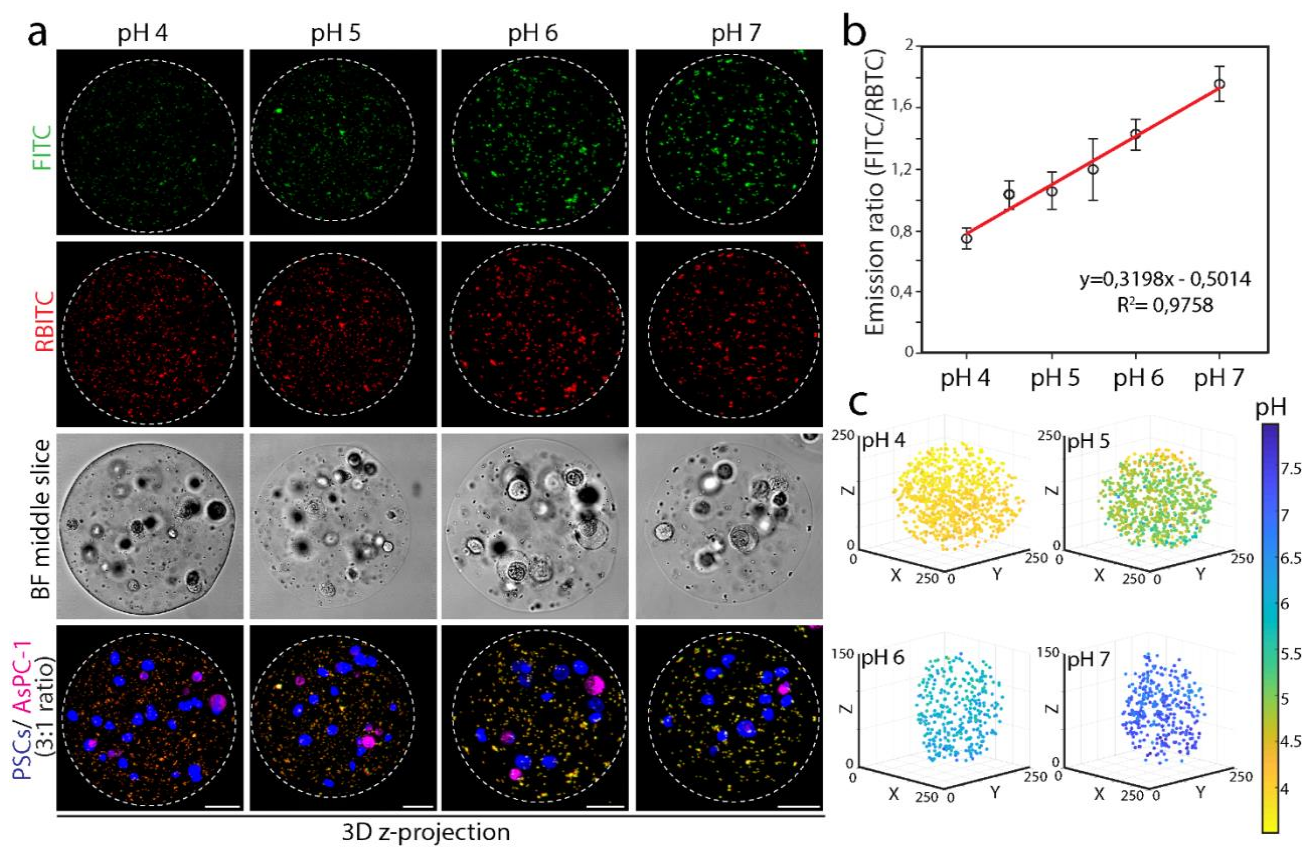

**Figure 4. Calibration of microparticles-based pH sensors in 3D spherical alginate hydrogels co-culture system**. **a**) Representative CLSM micrographs showing alginate hydrogels (maximum intensity projection) containing pH-sensor microparticles, PSCs and AsPC-1 cells exposed to different pH-adjusted cell media. FITC (green channel), RBITC (red channel), PSCs (Hoechst, blue channel), AsPC-1 (Deep Red, magenta channel) and bright field (BF, grey channel) are shown. Dashed lines indicate the edge of the hydrogels. Z-projections of 71 sections for pH 4; 67 sections for pH 5; 48 sections for pH 6; 47 sections for pH 7. Z-stack step size = 2.27 µm. Scale bars, 50 µm. **b**) Ratiometric calibration curve of microparticles-based pH sensors shown in a). Fluorescence intensity ratio of green (FITC) and red (RBITC) channels was calculated for each sensor microparticle (see methods) and their means for the tested pH values are reported in the graph. Data are means ± SEM. **c**) 3D scatter plots of microparticles-based pH sensors of the experiment shown in (**a**) and relative pH colormap. Fluorescence intensity ratio of green (FITC) and red (RBITC) channels was measured for each sensor and the corresponding pH value, based on the ratiometric calibration curve reported in (**b**), was extrapolated and converted in false colour.

## 2.5 Mapping extracellular pH over time and space at the single-cell level

To monitor the extracellular pH over time and space at single cell resolution, we generated mono- and co-cultures of 3D pH-sensor hydrogels of tumour and stromal pancreatic cells as described before (see Experimental section and scheme in **Figure 1d**). Specifically, PSCs and AsPC-1 cells were microencapsulated in alginate hydrogels containing pH sensors, either alone (**Figures 5a-d**) or in combination (**Figures 5e and f**) (3:1 ratio), and pH calibration curves (**Figures S7**) were set up before proceeding with the 3D time lapse CLSM imaging (**Figures 5a-f**). After imaging the whole hydrogel, cells with their related extracellular microenvironment (arbitrary chosen based on the mean diameter of cell type, see methods and **Figure S5**) were crop and a 3D sub-volume was generated for each cell type as described in Experimental section and then automatically analysed by the segmentation algorithm (**Figures 5b,d,f; Figures S6a-c**). The intensity ratio of each pH sensor was calculated over time and pH values, which were assigned according to the pH calibration formula, were converted in colour. Then, 3D maps were generated for each cell and the topographical visualization of dynamic changes of pH sensors around the cells was monitored over time and space (**Figures 5g-I** and **Figures S6d-f**). Within the same time point (**Figures 5g-i** and **Figures S6d-f**), the sensors surrounding each cells show different colours depending to their position along the x, y, z axis. These evidences reflect the heterogeneous distribution of proton pools in the extracellular environment, which can be visualized, for the first time, by the 3D pH-sensor scaffold we have generated.

Quantifications of the mean pH response of $SiO_2$-based pH sensors over time within the extracellular microenviroment surrounding single Aspc1 and PSC cells in mono- and co-cultures are shown in **Figures 5j-l**, respectively. Notably, tumour and stromal cells exhibited different spatio-temporal pH behaviours. In particular, 3D mono-cultures of Aspc1 (**Figures 5a,b; Figures S6a**) showed a decrease in pH from a value of 7.7 to a minimum of 6.8 after 4h to then rise again to 7.5 after 10h (**Figures 5g,j**; **Figures S6d**), while the extracellular pH of 3D mono-cultures of PSCs (**Figures 5c,d**; **Figures S6b**) started from 7.7 and increased over time to assess to a stationary value of around 8 (**Figures 5h,k**; **Figures S6e**). Conversely, we registered for AsPC-1 and PSCs in 3D co-cultures (**Figures 5e,f; Figures S6c**), a different extracellular pH as compared to mono-cultures. Importantly, the pH measured at time 0 was 6.6, which is the same value detected at the end of the experiment (at 10h), while small variations were observed in between which, however, never exceed the maximum pH value of 6.7 and the minimum pH value of 6.3, respectively (**Figures 5i,l**; **Figures S6f**). Most likely, the pH variations observed in co-cultures conditions could be due to cancer:stroma (AsPC-1:PSCs) cells interactions, which results to a metabolic reprogramming toward a glycolytic phenotype.(Chiarugi and Cirri 2016; Fiaschi et al. 2012)

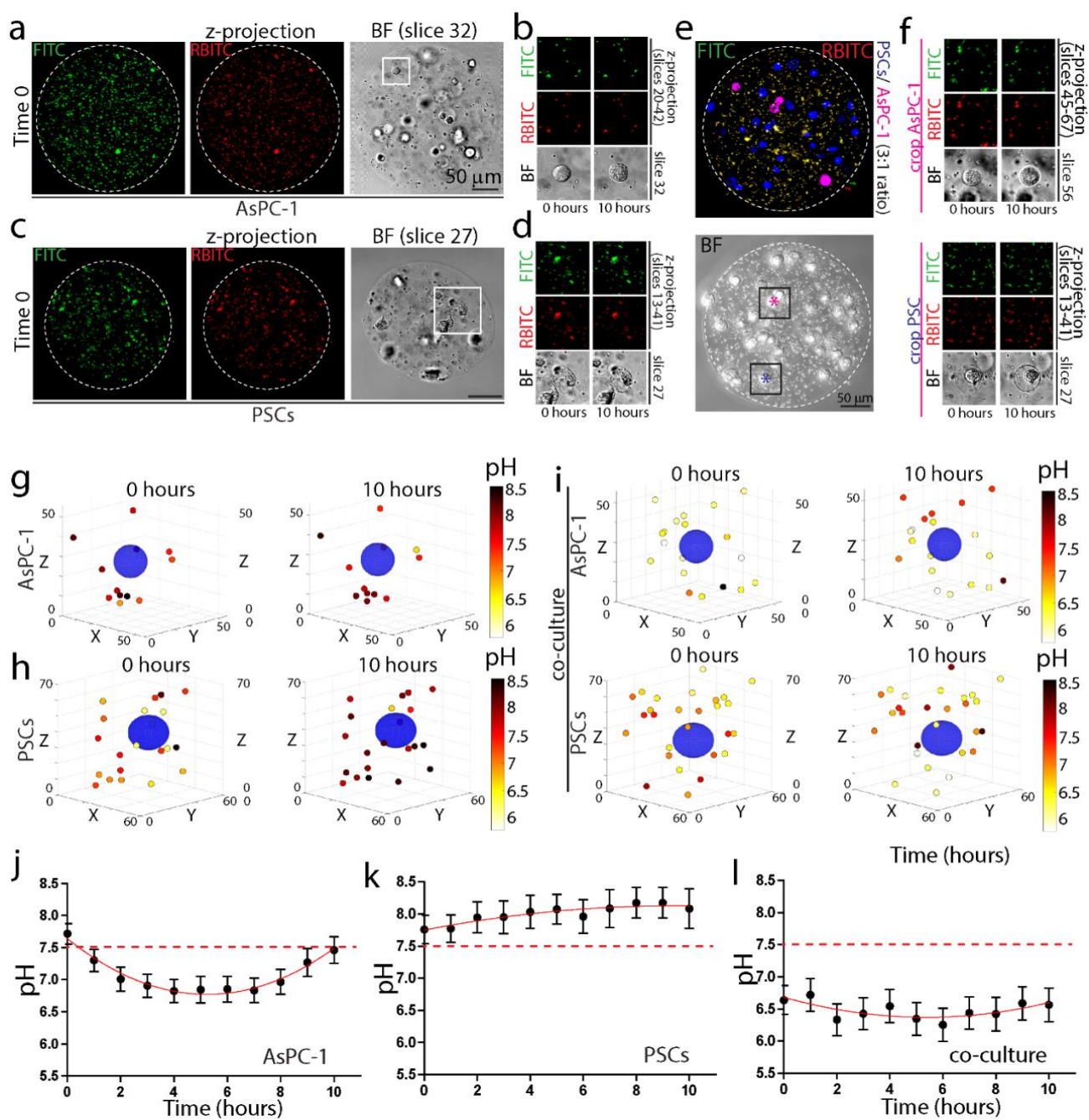

**Figure 5. Tumour-stroma extracellular pH evaluation over time and space.** Representative images of 3D time lapse of whole alginate hydrogel containing pH-sensor particles (FITC, green; RBITC, red) and ASPC1 cells in **a**, **b** (BF, grey; z-projection of 82 sections); PSCs in **c**, **d** (z-projection of 56 sections) and both (PSCs, Hoechst, blue; ASPC1, Deep Red, magenta) in **e**, **f** (z-projection of 75 sections). Dashed lines indicate the edge of the hydrogel. Z-stack step size = 2.27 µm. Scale bars, 50 µm. In **b** representative time-lapse images of T0 and T10 of selected cell in the boxed area in **a** (field of view in 50× 50 µm²; z-projection of 23 sections); in **d** representative time-lapse images of T0 and T10 of selected cell in the boxed area in **c** (field of view in 65× 65 µm²; z-projection of 29 sections); in **f** representative time-lapse images of T0 and T10 of a selected cancer cell (magenta asterisk in **e**, AsPC-1) and stromal cell (blue asterisk in **e**, PSC) in the black boxed areas in **e** (top panel field of view in 50× 50 µm². z-projection of 23 sections; bottom panel field of view in 65×65 µm². z-projection of 29 sections). In **g**, **h** and **i** are reported the 3D scatter plots of SiO₂-based pH sensors of the experiment shown in **b**, **d** and **f** respectively, with relative pH colormaps. The quantification of the experiments in **a**, **c** and **f** are reported in **j**, **k** and **l** respectively (see **Figure 2** and experimental section for details). The curve fitting (red line) was done using a second order polynomial function. Data are means ± SEM.

## 3. Conclusions

Here, we report the generation and characterization of sensitive and highly robust pH-sensor 3D scaffolds for localized and reversible pH sensing with micrometer-scale spatial resolution. By means of microencapsulation technology, $SiO_2$-based fluorescent ratiometric pH sensors, tumour and stromal pancreatic cells (alone or in combination) were embedded in 3D biocompatible alginate hydrogels and imaged under time lapse CLSM. After imaging of the whole hydrogel scaffold, followed by image processing and analysis, we were able to map the extracellular pH over time and space at single cell level, thus monitoring their mutual metabolic interplay in a 3D complex microenvironment. Dysregulated pH causing intracellular alkalization and extracellular acidosis is a hallmark of cancer metabolism (Persi et al. 2018), and it is associated with tumour development, progression and treatment resistance.(Robey et al. 2009; Webb et al. 2011) Therefore, such analyte-sensor 3D scaffolds represent a powerful tool for pH sensing, with unprecedented spatial and temporal resolution, in appropriate *in vitro* 3D tumour models,(Boedtkjer and Pedersen 2020) (Moldero et al. 2020) for deciphering the role of pH tumour heterogeneity and for assessing the response to anticancer treatments,(Anemone et al. 2019) key aspects in precision medicine.

The composition and the properties of the resulting 3D sensor platform can be tailored by using polymers with different functions (e.g., collagen, chitosan, Matrigel, and fibrinogen matrices) as well as by integrating sensors detecting multiple analytes (i.e., oxygen, calcium, etc). The presented system fully preserves both the growth of multiple cell types and the sensitivity of optical sensors to local analyte changes. Importantly, cells can be retrieved from alginate hydrogels by a simple de-gelling process(Yu et al. 2019) that does not require disaggregation of multi-cellular structures. The spatio-temporal pH maps are extracted by means of customized algorithms designed for fast processing of big CLSM data set (x, y, z, t) and automatic detection of sensors. Finally, the functional pH-sensor 3D scaffolds can be successfully fabricated on a large variety of supports making them highly compatible for clinical diagnostics, high-throughput bioassays and environmental monitoring.

## 4. Experimental Section

### Cell lines

Human pancreatic cancer cell line AsPC-1 (ATCC® CRL-1682™) and Panc1 (ATCC® CRL-1469™) were obtained from American Type Culture Collection (ATCC, Rockville, Md., USA) and cultured at 37 °C in a humidified 5% $CO_2$ incubator. AsPC-1 and Panc1 were grown in RPMI-1640 (Sigma-Merck KGaA, Darmstadt, Germany) supplemented with 10% FBS (Gibco), 2mM glutamine and 1% penicillin/streptomycin (Sigma-Merck KGaA, Darmstadt, Germany). Human immortalized pancreatic stellate cells (PSCs), kindly provided by Dr. Enza Lonardo, Institute of Genetics and Biophysics of Cnr, Naples, Italy were cultured in DMEM medium (Sigma-Merck KGaA, Darmstadt, Germany) supplemented with 10% FBS (Gibco), 2mM glutamine and 1% penicillin/streptomycin (Sigma-Merck KGaA, Darmstadt, Germany) at 37° C with 5% $CO_2$. Mycoplasma contamination was routinely tested by Mycoplasma PCR detection kit (Abm, Canada) and mycoplasma-free cells, cultured until 10 passages, were used for the experiments.

### Chemicals

Rhodamine B isothiocyanate (RBITC), Fluorescein 5(6)-isothiocyanate (FITC)ammonium hydroxide 28%, Calcein AM, propidium iodide, bisBenzimide H33342 trihydrochloride (Hoechst) and agarose low EEO were all purchased from Sigma-Aldrich, Darmstadt, Germany. Tetraethyl orthosilicate (TEOS) and (3-Aminopropyl) triethoxysilane (APTES) were purchased from Aldrich chemistry. Anhydrous ethanol from VWR, ethanol from Honeywell, Potassium Chloride (KCl) from Sigma life science. Fluorescent glucose analogue 2-NBDG (2-[N-(7-nitrobenz-2-oxa-1,3-diazol-4-yl) amino]-2-deoxy-D-glucose) and buffer solutions (pH 5; 6; 7; 8) were purchased from Invitrogen, Thermo fisher Scientific. Alginic acid sodium salt from brown algae (Sodium Alginate) were purchased from Sigma life science.

### Synthesis of SiO₂-based microparticle pH sensors

Silica seed particles were synthetized dissolving 23 mg of KCl in 9.45 ml of milliQ water, adding 95 ml of ethanol and lastly 6 ml of $NH_4OH$ 28% and 1.73 ml of TEOS to the reaction mixture, which was kept under stirring (240 rpm) at room temperature for 20 minutes. FITC-APTES and RBITC-APTES thiourea synthesis were conducted in tandem for 4 hours, under stirring at room temperature adding 3.3 mg of FITC and 4.2 mg of RBITC, respectively to 8.7µl APTES and 2 ml of anhydrous ethanol. Once the two syntheses were completed, crude thiourea products were united and added to 2 ml of anhydrous ethanol and 0.44 ml of TEOS. This solution was filled into a 10 ml syringe and injected into the $SiO_2$ microparticles flask with a flow rate of 0.05 ml/minute. The reaction mixture was kept under stirring (240 rpm) at room temperature for 24 hours. The final product was centrifuged at 1500 rpm for 5 minutes and the supernatant was discarded. The pellet was washed in 40 ml ethanol three times (1500 rpm, 5 minutes) and then in D.I. water other 3 times (1500 rpm, 5 minutes). From one synthesis we obtained 475 mg of sensors. The clean product was resuspended in 10 mL of ethanol (stock solution concentration 47.5 mg/mL) and stored in the dark at room temperature. Stock solution contained $5.38 \times 10^6$ sensors per millilitre, as measured by flow cytometry (CytoFLEX S, Beckman Coulter, USA). The average hydrodynamic diameter of the pH sensors resulted 1.521 ± 0.073 µm (PdI 0.099) as measured by Dynamic Light Scattering (DLS) (Zetasizer Nano ZS MALVERN) analyses (refractive index in water 1.458, absorption 0.010, 25°C, 3 minutes equilibrium time).(Gualberto J. Ojeda-Mendoza 2018)

**Characterization of SiO₂-based microparticle pH sensors**

The pH sensors were calibrated in the acidic-basic pH range (i.e., $5,0 - 5.5 - 6.0 - 6.5 - 7.0 - 7.5 - 8.0$) and the fluorescence analyses were performed by means of a fluorescence spectrophotometer (Cary Eclipse, VARIAN). To this aim, a stock solution of 47.5 mg/mL of SiO₂-based microparticles, dissolved in ethanol and kept in dark at room temperature, was properly dispersed by gentle vortex. Then, 100 μL of dissolved particles were added to 20 mL of Leibovitz's L-15 medium at room temperature. The solution was kept under stirring and its pH was continuously monitored by a pH meter (pH50, XS) while NaOH (1M) or HCl (1M) were added to the solution to adjust the pH to the desired value. Then, the sensors were analysed with the fluorimeter (488 nm for excitation of FITC, 500-700 nm for detection of FITC; 561 nm for excitation of RBITC, 565-700 nm for detection of RBITC) and the ratios between FITC and RBITC fluorescence intensities were calculated.

Similar procedure was performed for measuring the reversibility of the sensors and their aging, by evaluating a series of three cycles of switches between pH 7.0 and pH 5 at day 0 and after 7 days, respectively.

**Experimental steps for the production of pH-sensor scaffolds**

A lab-made microencapsulation system was setup and used for all experiments involving the generation of spherical alginate hydrogels (see Scheme in **Figure 1d**). The system is composed of a high voltage generator a syringe pump (World Precision Instruments, Model AL-4000) and collecting dish. The high voltage is generated by using a dc power source (iiBro® DC Power Supply, model i3005), to provide an input of 4.5 volts to a Boost Step up High Voltage Generator (DC 3.6V-6V to 20kV, TECNOIOT®). To convert the alternating high voltage generated by this step-up generator into DC voltage, a full-wave bridge rectifier is used. The full wave bridge rectifier was prepared using special high voltage diodes (5mA, 20kV). Finally, to further reduce the ripples, a ceramic capacitor (20 kV, 1 nF, C&G Semiconductor) is connected in parallel to the final output voltage. Considering the losses, the final output voltage is estimated to be ~ 10 K. This voltage was finally applied to the needle and the CaCl₂ solution.

For microencapsulation, cells were detached with 0.25% trypsin-EDTA and a pellet of $4 \times 10^6$ cells was resuspended in 250 μL of fresh medium. Similar procedure was performed for tumour and stromal cells co-encapsulation, with the only difference that pre-labelled stroma (Hoechst 33342, Sigma Aldrich) and cancer cells (CellTracker™ Deep Red, Invitrogen, ThermoFisher Scientific) were mixed in a 3:1 ratio. Then, 250 μL of 3% (w/v) of alginate (dissolved in deionized and sterile water) was added to the cell suspension together with 40 μL of pH sensors [$5.38 \times 10^6$ particles/mL of the pH sensors stock solution], before gentle mixing for 2 minutes. Notably, there are key parameters affecting the cell seeding density within the hydrogel that are **(i)** the number of cells per mL of prepolymer alginate solution, which is known to influence cell viability due to competing nutrient demands;(Gansau et al. 2018)**(ii)** the final concentration of prepolymer alginate solution, which influence the viscosity of the solution and the microgel's size;(Gansau et al. 2018) **(iii)** the mixing of the cells with the prepolymer alginate solution, which has to avoid bubble formation. The alginate solution containing cells and sensors was then collected with a syringe (BD Plastipak™ 1-mL Syringe) and placed into a syringe pump. Next, the positive cathode of the high-voltage generator was connected to the needle of the syringe (21 G blunt needle), while the negative cathode connected the voltage generator with a calcium chloride solution (100mM CaCl₂ and 0.4% w/v of Tween-20 dissolved in water) present in the petri dish, which was placed below the syringe tip at a distance of 3 cm. By activation of the syringe pump, at a constant flow rate of 0.05 mL min⁻¹ and in the presence of high voltage (V 4.5), drops of alginate solution were ejected from the needle. Such drops cross-linked as soon as they touched the calcium chloride solution, becoming gelled spheres with cells and sensors entrapped within. The alginate spheres were subsequently washed three times in complete medium, for 10 min each, and kept in the incubator under controlled temperature and 5% C0₂ before been processed for CLSM live cell imaging. The above-described experimental protocol was set up carefully and optimized to finally have 19.5 ± 2.7 µm (mean ± SEM) pH sensors per 50 µm$^3$ of alginate hydrogel.

**Cell viability assay**

PSCs cells were encapsulated with AsPC1 or Panc1 cells and unlabelled SiO$_2$ microparticles into spherical 3D alginate hydrogels, as described above, and the cell viability was evaluated until day 6. Briefly, after production the hydrogels were incubated in complete medium containing Calcein AM (Sigma-Aldrich, Darmstadt, Germany) and propidium iodide (PI, Sigma-Aldrich, Darmstadt, Germany), to a final concentration of 0.25 µM and 10 µM respectively. After 30 min of incubation, the hydrogels were washed with complete medium and incubated for 20 min with Hoechst 33342 (Sigma Aldrich, Darmstadt, Germany) to stain cell nuclei. Then, the hydrogels were plated into an 8-well chamber slides (IBIDI, Berlin, Germany) previously functionalized with 0.1 mg ml$^{-1}$ of poly-lysine (Sigma Aldrich, Darmstadt, Germany) in order to facilitate the adhesion of the hydrogels on the well, thus strongly reducing sample shift during time lapse with CLSM. Representative images were captured at days 1, 3, and 6 of culture using a CLSM Leica SP8 (Leica Microsystem, Manheim, Germany) at 20X magnification. The maximum projections of z-stack images were obtained using Image J software. Semi-automatic particle analysis was performed to estimate the percentage of dead cells. Briefly, the images were converted to "binary" (black and white) images. A threshold range was set to the dead cells, stained by PI, apart from the background. All pixels in the image with values under the threshold were automatically converted to black and all pixels with values above the threshold were converted to white. Similar procedure was performed for the Hoechst channel, thus the number of dead cells (PI positive) respect to the total number of cells (Hoechst positive) were quantified and express in percentage of the total.

**Glucose uptake in 3D alginate hydrogels**

The evaluation of cellular glucose uptake was performed by monitoring the cellular internalization of a fluorescent D-glucose analogue (2-NBDG) by both flow cytometry and fluorescent microscopy. After 1 day of culture, the cell-seeded hydrogels were serum starved for 12 hours in DMEM without glucose (Gibco, Life Technologies). Then, 2-NBDG was added to a final concentration of 100 µM and the uptake was monitored for 30, 60 and 180 minutes. At each time point, cells were extracted from alginate hydrogels by incubating them with EDTA 0.5 M, which dissolves alginate instantly (Yu et al. 2019). The samples were washed twice in PBS and then analysed by flow cytometry (CytoFLEX S, Beckman Coulter, USA). Approximately 20,000 gated events were acquired for each sample and analysed using CytExpert software. Dead cells and debris were excluded based upon forward scatter and side scatter measurements. In parallel, hydrogels containing AsPC1, Panc1 and PSC stained with Hoechst (nuclei), RCA I rhodamine and deep Red (plasma membrane) respectively, were exposed to 2-NBDG (100 µM) for 3 hours. After incubation, the hydrogels were washed three times with complete medium before been imaged with EVOS M7000 microscope (Invitrogen, Thermo fisher Scientific).

**Development of an algorithm for 4D image processing and analysis**

The algorithm was designed to automatically extract pH read-outs from optical sensors embedded in 3D structures. The code was written in GNU Octave (version 6.2.0).

In the following the computational pipeline (**Figure 3**) is described. The algorithm takes as input a CLSM time-lapse 3D sequence (x, y, z, t), with four fluorescence signals, two belonging to the sensors (FITC and RBITC) and one or two to the cell staining (Hoechst and Deep Red). pH is estimated in 7 main steps:

**a.** (i) Pre-processing of (x, y) cross-sectional images: single channel images were first converted to grayscale. Photon shot noise was reduced by median filtering of the image in two dimensions.

Image were binarized with Otsu's method (Otsu 1979). Then, morphological opening was performed to remove any small white noise in the image, and morphological closing with a disk-shaped structuring element to remove any small holes in the object. All connected components that had fewer than 8 pixels were removed and structures that were connected to the image border were suppressed.

(ii) The 2D images were segmented by a watershed transformation(Meyer 1994) and a distance transform(Calvin R. Maurer and Claims 2003) was used as segmentation function to split out the regions (**Figure 3a**). To avoid over segmentation a minima imposition procedure was implemented(Vincent 1993) first, tiny local minima were filtered and then, the distance transform was modified so that no minima occurred at the filtered-out locations. With watershed algorithm contiguous regions of interest were segmented into distinct objects, identifying single particles in the binarized FITC (*bwFITC*) and RBITC (*bwRBITC*) images and single cells in the Hoechst (*bwHoechst*) and Deep Red (*bwDR*) channels. *bwFITC* and *bwRBITC* images were passed as masks on the original FITC and RBITC images respectively (obtaining the *mFITC* and *mRBITC* images) to extract particle objects with original pixel intensity values, bringing the background to 0.

**b.** RBITC respect to FITC channel motion correction. During time-lapse CLSM imaging, temperature fluctuations and small vibrations of the microscopy stage might affect the imaging acquisition, thus causing sample drift (Jonkman et al. 2020). We observed these phenomena in our 3D time lapse data sets, in fact the sample imaged presented small movements in the (x-y) plane. Indeed, overlaying FITC and RBITC channel images, the positions of the pixels belonging to the sensors did not fit. The pixel movement was not systematic (i.e., the shift was not of the same magnitude and along the same direction for all the particles of the image). Specifically, shift was lower than 1 µm (part of the sensor in the FITC and RBITC channels was overlaid), and had random direction, (**Figure 3b**). Therefore, we first measured, iteratively for each sensor, the shift Euclidean distance $d_{RBITC\text{-}FITC}$ between the centroids in the FITC and RBITC channels. Then, we fixed the RBITC respect to FITC shift by moving of a distance $d_{RBITC\text{-}FITC}$, the pixels belonging to a specific sensor in the RBITC channel towards their position in the FITC channel, making the centroid in the RBITC channel match to the centroid in the FITC channel. Finally, in order to remove eventual sensors internalized by cells, particles in *bwFITC* and *bwRBITC* overlaying with the cell regions of *bwHoechst* and *bwDR* were excluded.

**c.** For each time points *t, bwFITC, mFITC*, shift-corrected *mRBITC, bwHoechst* and *bwDR* z-stack images were stored in three-dimensional matrices. Objects (cells and sensors) in 3D were then defined by directly connecting along the z-axis the 2D (x, y) binary images (**Figure 3c**). The basic steps for finding the connected components consisted in: (i) Finding an unlabelled pixel, *px*; (ii) Using a flood-fill operation to label all the pixels in the connected component containing *px (t*wo adjoining pixels were considered part of the same object if their faces, edges, or corners touched); (iii) Repeating the first two steps until all the pixels were labelled.

Then, outlier objects with cell nuclei or particles' sizes below a threshold were removed.

**d.** Starting from 3D binary images, unweighted coordinates (x, y, z) of the centroids for cells and sensors were extracted (**Figure 3d)** as the average of the *x, y, z* coordinates of all the pixels in the considered object.

**e.** Fluorescence intensity values ($I_{FITC}/I_{RBITC}$) (**Figure 3e**) were calculated as the pixel-by-pixel ratio between FITC and RBITC fluorescence intensities. In order to obtain $I_{FITC}/I_{RBITC}$ measured from a specific sensor, results were averaged along the pixels composing that sensor.

**f.** Then, starting from the overall spheroid data, the sensors found close to the tumour cells were selected (**Figure 3f**). Indeed, knowing centroids and sizes of cells and microparticles, sensors having Euclidean distance $d_{cs} < 2*R + 10$ $\mu m$ from each tumour cell were extracted, where $R = 16$ $\mu m$ is the cell ray. **g)** pH values were calculated from $I_{FITC}/I_{RBITC}$ intensities by passing data in a previously extracted $pH$- $I_{FITC}/I_{RBITC}$ calibration curve (**Figure 3g**). In fact, $I_{FITC}/I_{RBITC}$ values obtained from pH-fixed solutions were previously used to fit a curve and calculate the calibration formula connecting $I_{FITC}/I_{RBITC}$ values with pH.

Local pH measures were also aggregated into global indexes, by taking mean and standard deviation of the pH values over all the sensors belonging to a spheroid.

### Live cell imaging

After production, spherical alginate hydrogels were immobilized onto an 8-well chamber slide previously functionalized with 0.1 mg ml⁻¹ of poly-lysine. Then, samples intended for the evaluation of extracellular pH were gently washed three times in complete medium, for 5 min each and then incubated with pH 7.0 adjusted cell medium, while those intended for the setup of the ratiometric calibration curve were incubated with pH adjusted cell media (from pH 4.0 to pH 8.0). All the samples were left to equilibrate for 30 minutes before being imaged with CLSM. The sample were then mounted and analysed under a confocal microscope Leica TCS SP8 (Leica Microsystems GmbH, Wetzlar, Germany) (objective HC PL FLUOTAR 20x/0.50 DRY (0.5 NA)) under controlled temperature and $CO_2$. For time lapse experiments, z-stack images of the whole hydrogel were acquired at regular interval of 1 hour (488 nm for excitation, PMT3: 500-550 nm; 561 nm for excitation, PMT3: 570-620 nm (PMT TRAN ON); UV 405 for excitation, PMT1: 415-500 nm; 633 for excitation, PMT5: 640-750 nm; 1024x1024 pixels; zoom factor: 2).

Images were processed using ImageJ and analysed with the algorithm developed in **Figure 3** (see also experimental section). GNU Octave, Excel and GraphPad Prism version 5.0 software were used for data analyses and graphing.

### Statistical analysis

Error bars refer to either standard deviation (SD) or standard error of the mean (SEM) according to the different experiments and as indicated in the figure legends.

### Declaration of competing interest

The authors declare that they have no known competing financial interests or personal relationships that could have appeared to influence the work reported in this paper.

### Author Contributions

The manuscript was written through contributions of all authors. All authors have approved the final version of the manuscript.

### Acknowledgments

The authors gratefully acknowledge support from the European Research Council (ERC) under the European Union's Horizon 2020 research and innovation program ERC Starting Grant ''INTERCELLMED'' (contract number 759959), the My First AIRC Grant (MFAG-2019, contract number 22902), the PRIN-2022 (contract number 20205B2HZE_004) and the ''Tecnopolo per la medicina di precisione'' (TecnoMed Puglia) Regione Puglia: DGR n.2117 of 21/11/2018, CUP: B84I18000540002. We thank Dr. Carlo Giansante for performing TEM and Eliana D'Amone for technical support.

## Graphical Abstract

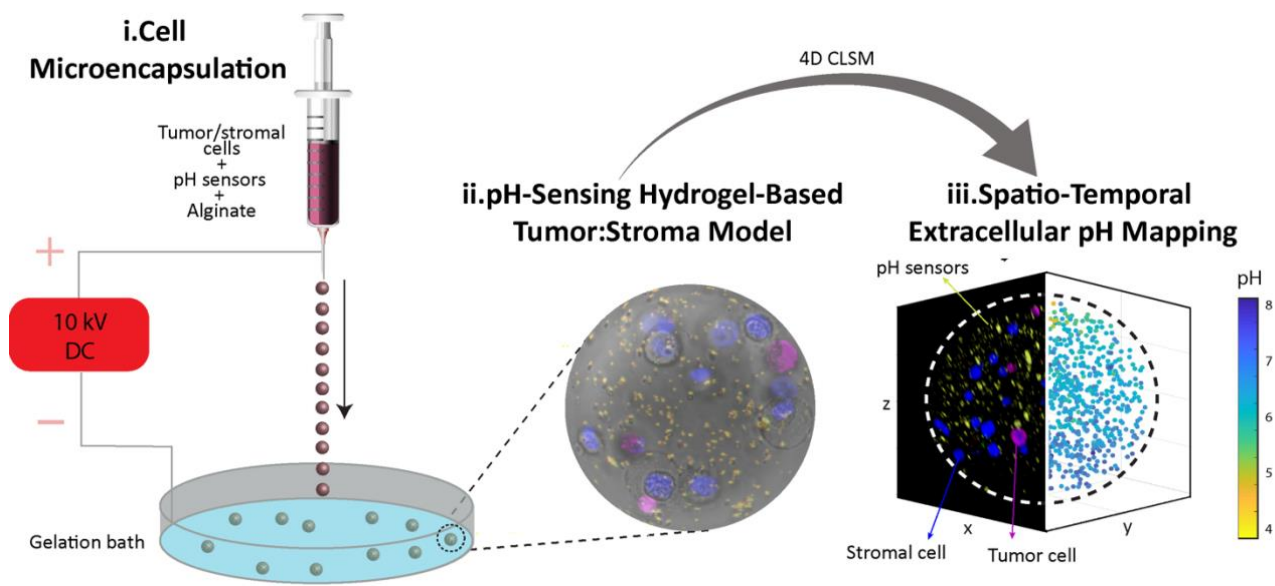